  \documentclass{gretsi}

   \usepackage[english,french]{babel}   
  \usepackage{times}			
 \usepackage{color}
\usepackage{epsfig}
\usepackage{amsmath}
\usepackage{amsfonts}
\usepackage{amsthm}
\usepackage{amssymb}
\usepackage{graphics}
\usepackage{graphicx}
\usepackage{float}
\usepackage{algorithm}
\usepackage{algorithmic}
\usepackage{svg}
\usepackage {epstopdf}
 
\title{On the latency of multi-level polar coded modulations}

\author{\coord{Vincent}{Corlay}{}}

\address{\affil{}{Mitsubishi Electric R\&D Centre Europe \\
         1 All. de Beaulieu, Rennes, France}}

\email{v.corlay@fr.merce.mee.com}

\frenchabstract{Un inconvénient communément admis du codage multiniveau, par rapport à une modulation codée avec entrelacement de bits,
est sa latence élevée : En effet, les niveaux doivent être décodés séquentiellement.
Dans cet article, nous considérons des codes polaires pour coder chaque niveau. Nous montrons que la latence de décodage du schéma multiniveau, utilisant le décodage par annulation successive avec liste pour chaque code polaire, n'est que de 1.5 fois celle d'un seul code polaire, quel que soit le rapport signal sur bruit et le nombre de niveaux.}

\englishabstract{A commonly assumed drawback of multi-level coding, compared to a bit-interleaved coded modulation,
is its high latency: Indeed, the levels must be decoded sequentially. 
In this paper, we consider polar codes to code each level. We show that the decoding time complexity of the multi-level scheme, using successive-cancellation list decoding for each polar code, is only 1.5 times the one of a single polar code, regardless of the signal-to-noise ratio and the number of levels. }

\begin{document}
\maketitle

\section{Introduction}

There exist two main techniques to build high-rate codes: bit-interleaved coded modulations (BICM) \cite{Caire1988} and multi-level coding (MLC) \cite{Imai1977}\cite{Seidl2013}\cite{Wachsmann1999}. 
On the one hand, a BICM cannot achieve optimal performance from an information theory perspective, even though the loss is sometimes insignificant. See e.g.,  \cite[Sec.~VI.A]{Bocherer2015} or \cite[Sec.~IV.A]{Smith2012}.
On the other hand, MLC is theoretically optimal. 
Nevertheless, BICM have often been preferred over MLC due to the following assumptions:
\begin{itemize}
\item Several binary codes have to be used with MLC: one for each level. This induces a high complexity.
\item The decoding latency of a MLC scheme is high because the levels (i.e., codes) have to be decoded sequentially. 
\item The paradigm of MLC is sometimes considered more complex.
\end{itemize}
As an example, we find the following sentences in the literature:
\begin{itemize}
\item In \cite[Sec.IV.A]{Smith2012}: ``Note that the multi-stage architecture introduces decoding latency to the higher levels ... clearly, the latency and memory issues can be eliminated simply by ignoring the conditioning (i.e., implementing a BICM)".
\item In the introduction of \cite{Barakatain2020}: ``MLC has often been avoided in optical communications because of the potentially high complexity induced by using separate bit-level codes and the negative impact that multi-stage decoding has on~latency".
\end{itemize}
Nevertheless, recent studies comparing both schemes tend to reconsider these assumptions. See for instance \cite{Barakatain2020}.

In this work, we add a favorable argument for MLC. 
We consider polar coding for each level with successive-cancellation list decoding. We show that the decoding latency of MLC combined with polar coding is similar to the one of a single polar code with rate $R=1/2$.


\section{Time complexity of a polar decoder}
To begin with, let us introduce the decoding time complexity (TC), denoted by $\mathfrak{C}$, to model the latency. It is defined as the number of time steps required to decode a codeword, where all the parallelizable instructions are performed in one clock cycle. 

\subsection{Polar codes}
As explained in \cite{Alamdar2011}, a polar code, introduced by Arikan in \cite{Arikan2009}, is defined by the following parameters: the block length $N=2^n$, the rate $R=K/N$, and an information set $\mathcal{A} \subseteq [N]$ of cardinality $K$, where $[N]=\{1,...,N\}$. The elements of $\mathcal{A}$ are the indices of the information bits and the one of $[N] \backslash \mathcal{A}$ the indices of the frozen bits.
The encoding for a polar code of length $N$ is performed via a modulo-2 matrix multiplication $x = u G_n$,
where $G_n$ is the generator matrix of the polar code (see \cite{Arikan2009}), $u = [u_1,...,u_N]$ is the input vector, and $x = [x_1,...,x_N]$ a codeword. Hence, $u_i$ is an information bit if $i \in \mathcal{A}$ and a frozen bit otherwise.

\subsection{Successive-cancellation (list) decoding}

We present the standard successive-cancellation (SC) decoder for polar codes. 
This decoder can be implemented as a message passing algorithm on a tree. 

Let $T_n$ denote a binary tree of depth $n$ and $v$ a node in the tree. 
The variables $v_p$, $v_l$, and $v_r$ refer to the parent node of $v$, and the left and right child node of $v$, respectively.

\begin{figure}[h]
\centering
\includegraphics[width=0.95\columnwidth]{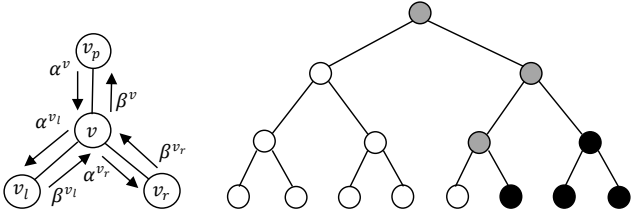} 
\caption{Left: Messages involved in the SC decoding algorithm. Right: Decoding tree $T_3$.}
\label{fig_decod}
\end{figure}

The SC decoding algorithm works as follows over $T_n$. 
Let $N_v$ represent the size of a message at node $v$.
Each node $v$ receives a message $\alpha^v=\{ \alpha^v_1, ..., \alpha^v_{N_v}\}$ from its parent node $v_p$, which contains $N_v$ logarithm likelihood ratio (LLR).
The messages $\alpha^{v_l}$ and $\alpha^{v_r}$, of length $N_v/2$, transmitted from $v$ to $v_l$ and $v$ to $v_r$, respectively, are computed as:
\small
\begin{align}
\begin{split}
&\alpha_i^{v_l} = 2 \text{arctanh}(\text{tanh}(\alpha_i^{v}/2) \text{tanh}(\alpha^v_{i+N_{v}/2})), \ 1\leq i \leq N_v/2,\\
&\alpha_i^{v_r} = \alpha^v_{i+N_v/2}+ (1-2\beta^{v_l}_i) \alpha^v_i, \ 1\leq i \leq N_v/2.
\end{split}
\end{align}
\normalsize
Bit estimates $\beta^v=\{\beta^v_1,..., \beta^v_{N_v}\}$ are passed from $v$ to its parent node $v_p$. 
The message $\beta^v$ is computed  from $\beta^{v_l}$ and $\beta^{v_r}$~as:
\small
\begin{align}
\begin{split}
\text{If } i \leq N_v/2, \ \beta_i^v = \beta_i^{v_l} \oplus \beta_i^{v_r}, \text{ if } i > N_v/2, \ \beta_i^v = \beta^{v_r}_{i-N_v/2}.
\end{split}
\end{align}
\normalsize
The messages are shown on Figure~\ref{fig_decod} (left).\\
The message $\alpha^{ro}$ of the root node of the tree is the LLR vector computed from the received vector $y$ (the output of the channel): \small $\alpha^{ro}_i=\text{log}(P(y_i|x_i=0)/P(y_i|x_i=1))$ \normalsize. At a leaf node $v$,  $\beta^v = \hat{u}_i = 0$ if $\alpha^v \geq 0$ and $\beta^v=1$ otherwise, where $1\leq i \leq N$ is the index of the leaf node $v$.

For short and moderate block-length polar codes, the performance of SC decoding can be improved via SC list (SCL) decoding \cite{Tal2015}: Instead of focusing on only one candidate, the $L$ most likely candidates are tracked. At an information leaf node, both the bit values 0 and 1 are considered. Consequently, the number of candidates (paths) doubles. Each path is associated with a path metric (PM) and the $L$ candidates with the smallest PM are kept. The path metric $\text{PM}_{i_j}$, corresponding to estimate $\hat{u}_{i_j}$ of the $i$-th bit at the $j$-th path, is computed as \cite{Bala2015}\cite{Ali2017}:
\small
$
\text{PM}_{i_j} = \sum_{k=1}^{i} \log(1+e^{- (1-2\hat{u}_{k_j}) \alpha_{k_j}}).
$
\normalsize

\vspace{-3mm}
\subsection{Decoding Rate-1, Rate-0, and Rep. nodes}

Let $\mathcal{I}_v$ be the set containing the indices of the leaf nodes that are descendants to $v$.
For a node $v$ in $T_n$, if $I_v \subseteq \mathcal{A}$, i.e., the leaf nodes that are descendants to $v$ are all information bits, we say that $v$ is a Rate-1 node.
Similarly, if $I_v \subseteq ([2^n] \backslash \mathcal{A})$,   i.e., the leaf nodes that are descendants to $v$ are all frozen bits, we say that $v$ is a Rate-0 node.
The decoding tree for $n=3$, as well as Rate-1 nodes (black) and Rate-0 nodes (white) for $\mathcal{A}=\{6,7,8\}$, are shown on Figure~\ref{fig_decod} (right).

It was shown in \cite{Alamdar2011} that with SC decoding, Rate-0 and Rate-1 nodes can be efficiently computed without visiting the subtree rooted at the given node.
For a Rate-0 node $v$, the components of $\beta^v$ are  immediately set to 0 (assuming that the frozen bits have value 0). 
The TC is 1. 
For a Rate-1 node (Lemma~1 in \cite{Alamdar2011}) $\beta^v = 0$ if $\alpha^v \geq 0$ and $\beta^v = 0$ otherwise.
The TC is also 1.

These results on Rate-0 and Rate-1 nodes were extended to the case of SCL decoding in \cite{Ali2016}\cite{Ali2017}.
For a Rate-0 node, no new path is created. The PM of the $L$ existing paths $j$ of node $v$ are updated as:
\small
$
\text{PM}_{v_j} = \sum_{k=1}^{N_v} \log(1+e^{-\alpha_{k_j}}).
$
\normalsize
Hence, the TC is the cost of adding $N_v$ numbers, i.e., 
\small $\mathfrak{C}_\text{Rate-0} (v)= \log_2 N_v$\normalsize.

For a Rate-1 node, new paths are created. The PM of the $j$-th path of node $v$ is computed as:
\small
$
\text{PM}_{v_j} = \sum_{k=1}^{N_v} \log(1+e^{- (1-2\beta_{k_j}) \alpha_{k_j}}).
$
\normalsize
Theorem~1 in \cite{Ali2017} proves that only the $L$ first $\alpha_{k_j}$ with the lowest value need to be considered for path splitting. Then, for all the surviving paths,  for $N_v-L\leq k \leq N_v$, $\beta_{k_j} = 0$ if $\alpha_{k_j} \geq 0$ and $\beta_{k_j} =1$ otherwise. 
Consequently, the TC is 
\small
$\mathfrak{C}_{\text{Rate-1}}(v,L)=\min(N_v,L)$.
\normalsize

In addition to these two categories of nodes, we also consider Repetition nodes, where only the rightmost
leaf is an information bit. It is shown in \cite{Ali2016} that the TC of a Repetition node $v$ is
\small $\mathfrak{C}_{\text{Rep}}(v) = 1 + \log_2 N_v.$\normalsize


\subsection{Decoding time complexity of polar codes}

We make the following assumptions to compute the TC of SCL decoding, summarized in Table~\ref{table_complex}. 
They are similar to the one considered in \cite[Sec. V]{Alamdar2011}.

Rate-1, Rate-0, and Repetition nodes are discussed in the previous subsection.

 For a non-leaf (standard) node $v$:
One clock cycle is used to calculate $\alpha^{v_l}$ (once $\alpha^v$ is received). One clock cycle is used to calculate $\alpha^{v_r}$ (once $\beta^{v_l}$ is received).
 One clock cycle is used to compute $\beta^v$ (once $\beta^{v_r}$ is received).
 The time to wait the messages from the child nodes is $\mathfrak{C}(v_l) + \mathfrak{C}(v_r)$.

 For a leaf node $v$: 
 For a frozen-bit leaf node, there is no path splitting. One clock cycle is used to set $\beta^v$ to 0 and to update the PM.
 For an information-bit leaf node, there is a path splitting. One needs to compute the values of the new paths, and to sort and select the surviving paths (and compute $\beta^v)$. For simplicity, we assume that it is done in one clock cycle as for a frozen-bit leaf node.

\begin{table}[h]
\small
\begin{center}
\begin{tabular}{|c|c|}
 \hline
    &  \\
Rate-1 node & $\mathfrak{C}_{\text{Rate-1}}(v,L) = \min (N_v,L)$  \\
  \hline
   &     \\
Rate-0 node  & $\mathfrak{C}_{\text{Rate-0}}(v) =  \log_2 N_v $ \\
  \hline
   &  \\
Repetition node  & $\mathfrak{C}_{\text{Rep}}(v) =  1+\log_2 N_v $ \\
  \hline
   &  \\
Standard node  & $ \mathfrak{C}_{\text{Standard}}(v)  = 3 + \mathfrak{C}(v_l) + \mathfrak{C}(v_r)$  \\
\hline
   &  \\
Leaf node  & $ \mathfrak{C}_{\text{Leaf}}(v)  = 1$  \\
\hline
\end{tabular}
\end{center}
\vspace{-3mm}
\caption{Assumptions on the TC of each category of nodes.}
\label{table_complex}
\normalsize
\end{table}
For a given information set $\mathcal{A}$, Algorithm~\ref{main_alg_lat} enables to compute the TC of the polar code. 

\begin{algorithm}
\caption{TC of SCL decoding of a polar code.}
\label{main_alg_lat}
\textbf{Function} TC($v, L,\mathcal{A}$) \\
//The first call of the function should be done with the root node $v_{ro}$ of $T_n$. 
\begin{algorithmic}[1]
\IF{$v$ is a Leaf node}
\STATE TC = $\mathfrak{C}_{\text{Leaf}}(v)$
\ELSIF{$v$ is a Rate-1 node (i.e., if $\mathcal{I}_v \subseteq \mathcal{A}$)}
\STATE TC = $\mathfrak{C}_{\text{Rate-1}}(v)$
\ELSIF{$v$ is a Rate-0 node (i.e., if $\mathcal{I}_v \subseteq ([N] \backslash \mathcal{A}$) ) }
\STATE TC = $\mathfrak{C}_{\text{Rate-0}}(v,L)$
\ELSIF{$v$ is a Repetition node}
\STATE TC = $\mathfrak{C}_{\text{Rep}}(v)$
\ELSE
\STATE TC = 3 + TC($v_l,L,\mathcal{A}$) + TC($v_r,L,\mathcal{A}$) \\ //left child node + right child node
\ENDIF
\STATE \textbf{Return} TC.
\end{algorithmic}
\end{algorithm}

\begin{figure}[h]
\centering
\includegraphics[width=0.95\columnwidth]{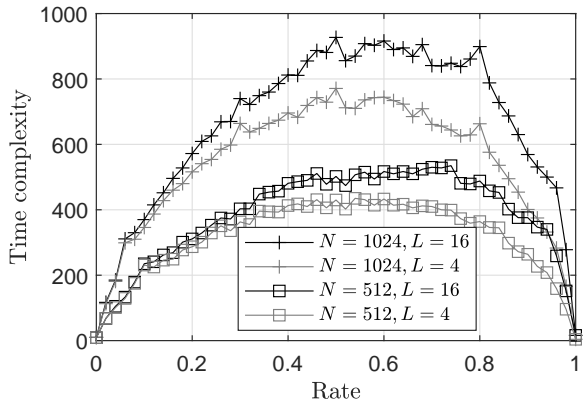}
\caption{TC of the SCL decoder as a function of the rate of the polar code with $N=512,1024$ and $L=4,16$. The codes are designed for the Gaussian channel via density evolution. 
}
\label{fig_complex_PC}
\end{figure}

We assess the decoding TC for polar codes designed for the Gaussian channel as follows.
For rates $0 \leq R \leq 1$, we find the set $\mathcal{A}$ (via density evolution\footnote{No optimization on the location of the information bits is performed to reduce the complexity.}), and we apply Algorithm~\Ref{main_alg_lat}.
The results are shown on Figure~\ref{fig_complex_PC}. As expected, the worst-case rate is around 0.5, where $\mathcal{A}$ induces a structure without many interesting Rate-1 and Rate-0 nodes. Significant improvements are observed for lower and higher rates.


\section{Time complexity for polar MLC}
\label{app_multilevelcoding}
\subsection{Multi-level coding}

\begin{figure}[h]
\centering
\includegraphics[width=0.7\columnwidth]{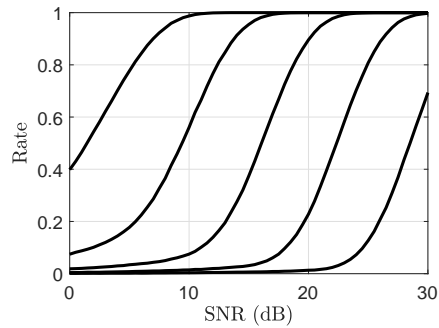}
\vspace{-3mm}
\caption{Rate of each level for a 32-ASK constellation.}
\label{fig_rate_32ASK}
\end{figure}

The constellation $\mathcal{X}$ considered in this paper is a $M$-amplitude-shift keying (ASK) constellation. The symbols of a $M$-ASK constellation, where $M=2^m$, are \small $\mathcal{X}=\{-2^m+1,..,-3, -1,$\\$+1,+3,\ldots,+2^m-1\}$\normalsize. Hence, $m$ is the number of bit levels.
Using the chain rule, the mutual information between the input of the channel $X$ and the output $Y$ can be expressed as
\small
$
I(X;Y) = I(B_1,B_2,...,B_m;Y)= \sum_{i=1}^m I(B_i;Y|B_1,...,B_{i-1}),
$
\normalsize
where $B_i$ denotes the random variable corresponding to the $i$-th bit of the labelling considered.
One bit level refers to the channel described by $I(B_i;Y|B_1,...,B_{i-1})$. When a binary code is used to transmit information over this $i$-th level the coding rate should be chosen to match\footnote{In practice, a back-off which depends on the code used is applied.} $\small I(B_i;Y|B_1,...,B_{i-1}) \normalsize$. Figure~\ref{fig_rate_32ASK} shows the rates of the five bit levels of a 32-ASK constellation with natural labelling (and with a uniform distribution of the symbols\footnote{If shaping is used, e.g., as in \cite{Corlay2022}, the positions of the curves are slightly shifted but this does not change the result in terms of TC.}) as a function of the signal-to-noise ratio\footnote{Defined as $E_s/\sigma^2$, where $\sigma^2$ the variance of the noise.} (SNR). 

Note that if the rate is close to 1, the information does not need to be coded as the mutual information equals the entropy.
If the level is coded, the code contains only information bits and therefore is decoded as a (SC) Rate-1 node.
If the rate is close to 0, then the polar code contains only frozen bits. It is decoded as a (SC) Rate-0 node. 
We observe that for any SNR, we have either:
\begin{itemize}
\item One level with a low rate but greater than 0, one with a high rate but smaller than 1, and all others close to 0 or~1.
\item One level with a rate close to 0.5 and all others close to 0 or 1.
\end{itemize}
We recall that multi-stage decoding involves using the results of the lower levels to decode the higher levels.
Hence, the TC of the MLC scheme is the sum of the TC of each level. 
Consequently, with the above observation we expect the TC to remain stable with the SNR and not significantly higher than the one of a single polar code with rate $R=0.5$.

\vspace{-3mm}
\subsection{Time complexity of multi-level polar coding}

We use Algorithm~\ref{main_alg_lat_2} to compute the decoding TC of the MLC scheme. 
It consists in first finding the information set $\mathcal{A}_i$ for each level $i$, then computing the TC of the corresponding polar code with Algorithm~\ref{main_alg_lat}, and finally summing the results of each level (again, we recall that the levels are decoded sequentially). Moreover, if the rate of a level is close to 0 or 1 (within $\epsilon$, see the algorithm), we set it to 0 or 1, as commonly done with MLC schemes. 

\begin{algorithm}[h]
\caption{Decoding TC of a multi-level polar coded modulation.}
\label{main_alg_lat_2}
\textbf{Input:} SNR, $\epsilon$, $m$,$L$. \\   //We take $\epsilon = 0.01$.
\begin{algorithmic}[1]
\STATE Compute the rate of each level \small $I(B_i;Y|B_1,...,B_{i-1})$\normalsize (which depends on the SNR).
\STATE Find the set of information bits $\mathcal{A}_i$ corresponding to \small $I(B_i;Y|B_1,...,B_{i-1})$ \normalsize for each level.
\STATE Set $\mathfrak{C}=0$.
\FOR{$1\leq i \leq m$}
\IF{$R_i< \epsilon$}
\STATE $\mathfrak{C}=\mathfrak{C}+1$.
\ELSIF{$R_i>1-\epsilon$}
\STATE $\mathfrak{C}=\mathfrak{C}+1$.
\ELSE
\STATE $\mathfrak{C} = \mathfrak{C} +$TC($v_{ro},L,\mathcal{A}_i$) //Algorithm~\ref{main_alg_lat}
\ENDIF
\ENDFOR
\STATE \textbf{Return} $\mathfrak{C}$.
\end{algorithmic}
\end{algorithm}

\begin{figure}[h]
\centering
\includegraphics[width=0.9\columnwidth]{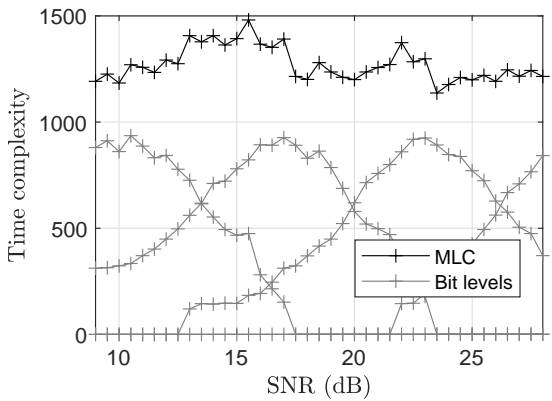}
\caption{Decoding TC for the MLC scheme with block length $N=1024$ and $L=16$. 
The black curve is obtained by summing the values of the grey curves for a given SNR.
}
\label{fig_result_multi_level}
\end{figure}

The result for a 32-ASK constellation with natural labelling is shown on Figure~\ref{fig_result_multi_level}. As expected, the decoding TC does not strongly depend on the SNR on thus on the data rate. Moreover, we see that the TC of the MLC scheme is approximately 1.5 times the worst-case complexity of a single polar code. 

\vspace{-4mm}
\section{Conclusions}
In this paper, we investigated the decoding TC of a MLC scheme where polar codes are used to code each bit level.
On the one hand, the TC of the optimized SCL decoder for a single polar code varies significantly with the rate. In particular, it is very low if the rate of the polar code is close to 1 or 0. On the other hand, the rates of each level, if natural labelling with a $M$-ASK  constellation is used, are all close to 0 or 1, with the exception of at most two levels. This holds even for large $M$ and regardless of the SNR. Consequently, the decoding latency of multi-level polar coded modulations is only slightly higher than the one of a single polar code.

\end{document}